\documentclass{article}

\usepackage{tikz}
\usepackage{subfig}
\usepackage{svg}
\usepackage{arxiv}
\usepackage{amsmath}
\usepackage[utf8]{inputenc} %
\usepackage[T1]{fontenc}    %
\usepackage{hyperref}       %
\usepackage{url}            %
\usepackage{booktabs}       %
\usepackage{amsfonts}       %
\usepackage{nicefrac}       %
\usepackage{microtype}      %
\usepackage{lipsum}		%
\usepackage{graphicx}
\usepackage{natbib}
\usepackage{doi}
\usepackage[margin=2cm]{caption}
\usepackage{float}
\usepackage[utf8]{inputenc}
\DeclareUnicodeCharacter{200A}{}

\title{Earthquake Magnitude and b value prediction model using Extreme Learning Machine}

\author{ \href{https://orcid.org/0000-0002-1318-7447}{\includegraphics[scale=0.06]{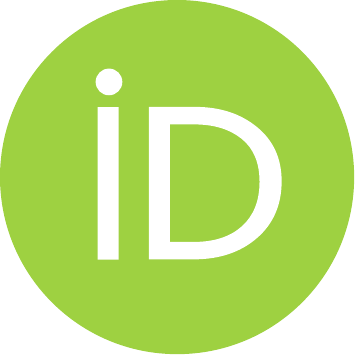}\hspace{1mm}Gunbir Singh Baveja}\thanks{Research was not conducted under the affiliation of University of British Columbia, Vancouver.} \\
	University of British Columbia\\
	Vancouver, BC V6T 1Z4 \\
	\texttt{gbaveja@student.ubc.ca} \\
	\And
	\href{https://orcid.org/0000-0003-0819-8851}{\includegraphics[scale=0.06]{orcid.pdf}\hspace{1mm}Jaspreet Singh} \\
	Department of Computer Science and Engineering\\
	GD Goenka University\\
	Sohna Rural, India \\
	\texttt{jaspreet.singh@ggdu.org} \\
}

\date{November 30, 2021}

\renewcommand{\undertitle}{Technical Report}

\hypersetup{
pdftitle={A template for the arxiv style},
pdfsubject={q-bio.NC, q-bio.QM},
pdfauthor={Gunbir Singh Baveja, Jaspreet Singh},
pdfkeywords={First keyword, Second keyword, More},
}

\begin{document}
\maketitle

\begin{abstract}
	Earthquake Prediction has been a challenging research area for many decades, where the future occurrence
of this highly uncertain calamity is predicted. In this paper, several parametric and non-parametric features
were calculated, where the non-parametric features were calculated using the parametric features. 8
seismic features were calculated using Gutenberg-Richter law, total recurrence time, seismic energy release.
Additionally, criterions such as Maximum Relevance and Maximum Redundancy were applied to choose the
pertinent features. These features along with others were used as input for an Extreme Learning Machine
(ELM) Regression Model. Magnitude and Time data of 5 decades from the Assam-Guwahati region were
used to create this model for magnitude prediction. The Testing Accuracy and Testing Speed were computed
taking Root Mean Squared Error (RMSE) as the parameter for evaluating the model. As confirmed by the
results, ELM shows better scalability with much faster Training and Testing Speed (up to thousand times
faster) than traditional Support Vector Machines. The Testing RMSE (Root Mean Squared Error) came out to
be. To further test the model’s robustness, magnitude-time data from California was used to- calculate the
seismic indicators, fed into neural network (ELM) and tested on the Assam-Guwahati region. The model
proves to be successful and can be implemented in early warning systems as it continues to be a major part
of Disaster Response and Management.
\end{abstract}


\section{Introduction}
Earthquake is one of the most destructive
and the deadliest natural disaster that has caused
thousands of deaths, and millions if not billions of
dollars in property loss. No part of the world is
immune to earthquakes. The developing
countries, in particular, are the most affected
because emergency response services may not be
available even in stressful times. A reliable early
warning system may potentially be able to save
lives and land. Since the biggest and deadly
earthquakes started occurring near the 1950s,
primitive prediction methods were thought as
necessities. But by the 1990s, continuing failure
led to many questions whether it was even
possible to foretell the Time-Location of
Earthquakes.

\begin{figure}
	\centering\includegraphics[scale=0.4]{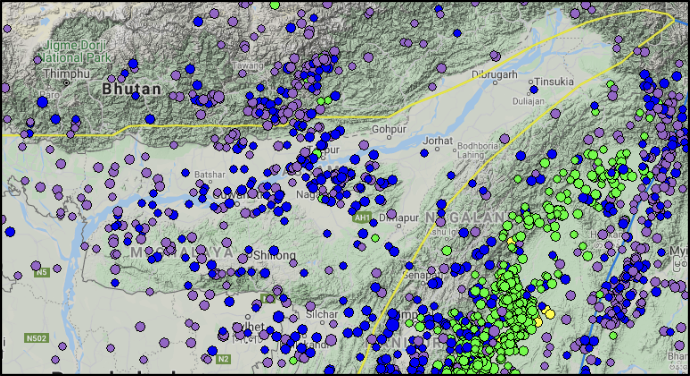}
	\caption{All the earthquakes ($1653$) in the north-eastern region of India (Assam-Guwahati) between $1933-2020$.}
	\label{fig:fig1}
\end{figure}

Fortunately, with the emergence of
modern computer science based intelligent
algorithms, it is relatively easy to predict or
classify data which has a definite pattern.
Significant results have been attained in
different fields of study such as Flood
Forecasting \citep{sagnik}, Weather
Analysis and Forecasting \citep{pmishra},
and disease diagnosis \citep{cosmag}.
Machine Learning and seismology can be linked
together to produce considerable results.

Assam-Guwahati region is the most
earthquake prone region of India due to
subducting Indian plate under the Eurasian plate.
This region has been experiencing earthquakes of
significant magnitude at moderate depth. A
polygon shaped Assam region was selected for
calculation of seismic features. Machine Learning
and Computational Intelligence has led to
paradigm shift in the methods of predictions and
determination of earthquakes. An ideal
earthquake predictor must yield the Magnitude,
Time, Energy release and Location of the
earthquake. Although our prediction model isn’t
perfect, it is indeed a huge step forward in
Earthquake research. 

The core idea of this work is
to predict earthquakes with magnitude $4.5+$ and
also predict the number of days in between
successive earthquakes. The mathematically
calculated seismic features were used as an input
to the SLFN (Single Layer Feed-Forward Network).
The prediction results of the neural network and
SVM were compared and discussed in this paper.

\subsection{Tectonics of Assam region}

Assam is one of the most seismically active
regions in India with events occurring at shallow
depth (0-70km). The region was geologically
formed due to collision of Eurasian and Indian
Plate during Eocene. The seismic parameters are
mathematically calculated from a catalogue, so
the catalogue should be complete above the cut-
off magnitude. Here, the cut-off magnitude refers
to the earthquake magnitude, below which
seismic events are not considered for parameter
calculations ($4.0$ in our case). See Section \ref{sec:param}.

\begin{figure}[H]
	\centering\includegraphics[scale=0.17]{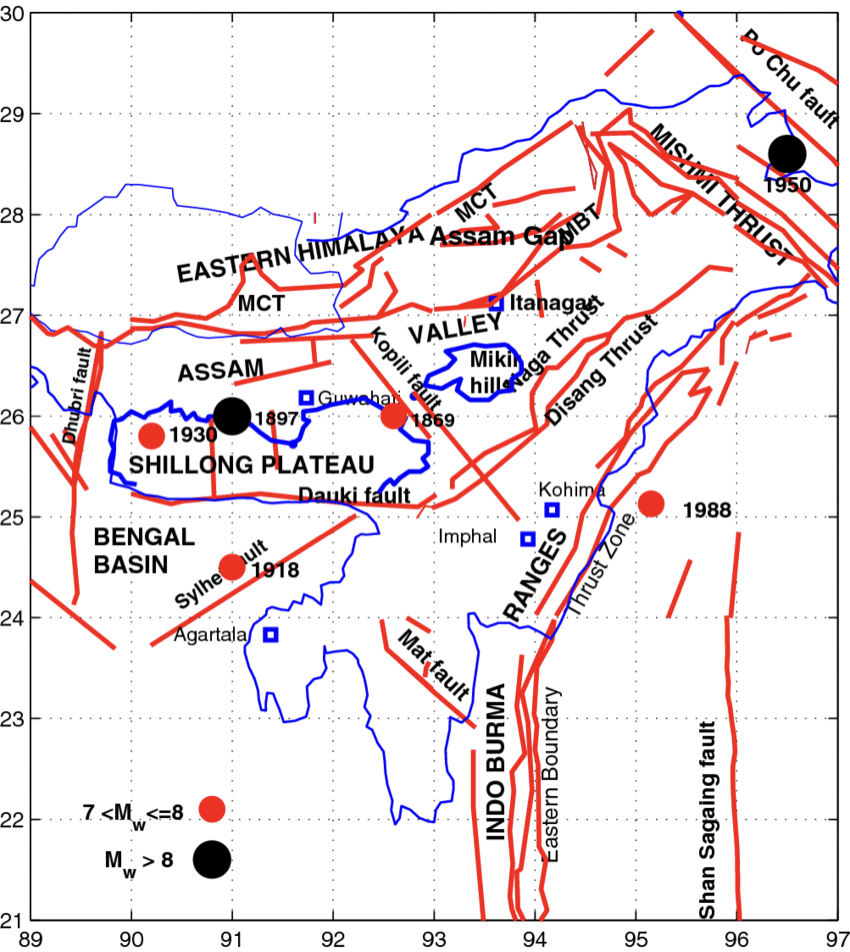}
	\caption{The Tectonic plates of the North-Eastern region of India (Assam-Guwahati region)}
	\label{fig:fig2}
\end{figure}

\section{Literature Review}
\label{sec:lit}
Various studies have been carried out by
researchers over earthquake occurrences and
predictions leading to various conclusions. The
Gutenberg-Richter mathematical model is one
such example where the relationship between
earthquake magnitude and frequency was
calculated; the relationship is analysed and used
to predict distribution over time. \citep{petersen} carried out research under the umbrella of
California Geological Survey (CGS) and proposed a
time-independent model showing that probability
of earthquake occurrence follows poisson’s
distribution model.

Several approaches have been proposed in the literature for using artificial neural networks (ANNs) to predict earthquakes based on various seismic precursors. For example, \citep{negarastani} used a backpropagation neural network (BPNN) to identify abnormal changes in soil radon concentration, which can be induced by earthquakes, by differentiating them from normal environmental variations. \citep{rbfliu} employed an ensemble of radial basis function (RBF) neural networks to forecast earthquakes in China using historical magnitude data as input. \citep{ikram} presented an expert system-based method for earthquake prediction that involves dividing the globe into four quadrants, using historic earthquake data as input and applying predicate logic and association rules to make predictions for each quadrant over a 24-hour period.

In their paper, \citep{kmasim} used four machine learning techniques, including a pattern recognition neural network, a recurrent neural network, a random forest, and a linear programming boost ensemble classifier, to predict earthquake magnitudes in the Hindukush region using a temporal sequence of past seismic activity. Earthquake precursors are phenomena that occur before a main shock and are causally linked to it, rather than simply occurring before it in time \citep{habberman}. These precursors can be based on continuous observations of various physical parameters such as seismic wave velocity, gravity, resistivity, and electricity \citep{nuannin}. For example, \citep{lu} found that drops in underground water levels and changes in resistivity recorded by geoelectric stations within $180$ km of the epicenter preceded the $M=7.8$ Tangshan earthquake. 

Other proposed precursors include changes in seismicity rates, source parameters of earthquakes, and frequency-magnitude distributions (FMD) (\citep{nuannin}; \citep{enescu}; \citep{huang}; \citep{Monterroso2003SpatialVO}; \citep{nuannin}; \citep{schoro}; \citep{wiemer}; \citep{wyss}; \citep{martir}). Seismic quiescence, or periods of significantly reduced seismicity, has also been suggested as a potential precursor \citep{katsumata}.
\begin{figure}[H]
	\centering\includegraphics[scale=0.4]{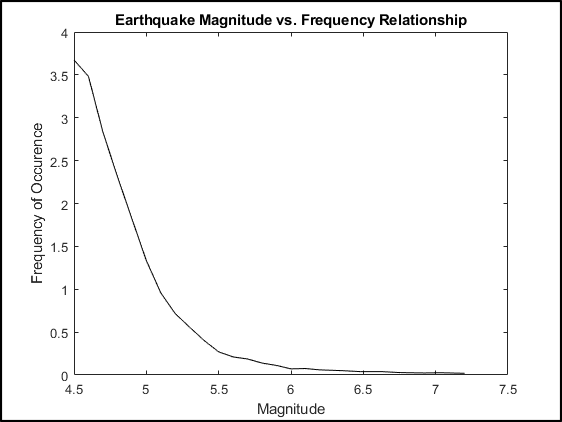}
	\caption{This graph proves the Inverse Square Law between the Magnitude and Frequency of the Earthquakes}
	\label{fig:fig5}
\end{figure}

\section{Seismic Parameters}\label{sec:param}
This study is carried out by using the eight seismic
indicators, which are basically meant
to represent the seismic state and potential of the
ground. This section contains the overview of all
the parameters and their calculation. One of the
parameters is the Time T, which is the time span
over the last n number of events and n in our case
is 100 and t represents the time of earthquake
occurrence.
\begin{equation}
T=t_n-t_1\end{equation}

\subsection{Mean Magnitude}
Time T represents the frequency of foreshocks
before the month under consideration. The
second seismic indicator considered is the mean
magnitude of the last n events. It relates to the
magnitudes of foreshocks, since the magnitude M
of seismic activity increases before a larger
earthquake. 
\begin{equation}
M_\text{mean}=\frac{\sum_i M}{n}.
\end{equation}
\subsection{Seismic Energy}
The rate of square root of seismic energy release dE is another seismic indicator that can be related to seismic activity through the phenomenon of seismic quiescence. Seismic energy releases gradually from fault lines through low-magnitude seismic events but if this phenomenon gets disturbed, it may lead to a major seismic event. The equation for square root of seismic energy released is given below:
\begin{equation}\mathrm{d}E^{\frac{1}{2}}=\frac{\sum\left(10^{10.8+1.5M}\right)^\frac{1}{2}}{T}\end{equation}

\subsection{\texorpdfstring{$a$}{a} and \texorpdfstring{$b$}{b} value}
The Frequency-Magnitude Distribution describes the number of earthquakes occurring in a given region as a function of their magnitude M as:
\begin{equation}\log N_i =a-bM_i,\footnote{This is the equation for Gutenberg-Richter law, describing the relationship between Magnitude and frequency of earthquakes}\label{gr_law}\end{equation}
where $N$ is the cumulative number of earthquakes with magnitude equal to or larger than $M$, and $a$ and $b$ are real constants that may vary in space and time.

The parameter $a$ characterizes the general level of seismicity in a given area during the study period, i.e., the higher the $a$ value, the higher the seismicity \citep{nuannin}. The parameter $b$ is believed to depend on the stress regime and tectonic character of the region \citep{bhatt}.

The $a$ and $b$ values are calculated numerically through two different methods. In earthquake prediction study for North-Eastern India, linear least square regression analysis based method is proposed.

\begin{align}
b_\text{lsq} &= 
\frac{\left(n\sum M_i\log N_i-\sum M_i\sum\log N_i\right)}{\left(\sum M_i\right)^2-n\sum M_i^2}\tag{5}\label{b_lsq} \\
b_\text{mlk}&=\frac{\log_{10} e}{\text{mean}(M)-\min(M)}\tag{6}\label{b_mlk}\\
a_\text{lsq}&=\sum\frac{\left(\log_{10} N_i+b_\text{lsq}M_i\right)}{n}\tag{7}\label{a_lsq}\\
a_\text{mlk}&=\log_{10} N+b_\text{mlk}\cdot\min(M).\tag{8}\label{a_mlk}
\end{align}
\ref{b_lsq} and \ref{a_lsq} represent the 
linear least square regression while \ref{b_mlk} and \ref{a_mlk} show the maximum 
likelihood method for calculation of $a$ and $b$ values.

\begin{figure}
	\centering\includegraphics[scale=0.5]{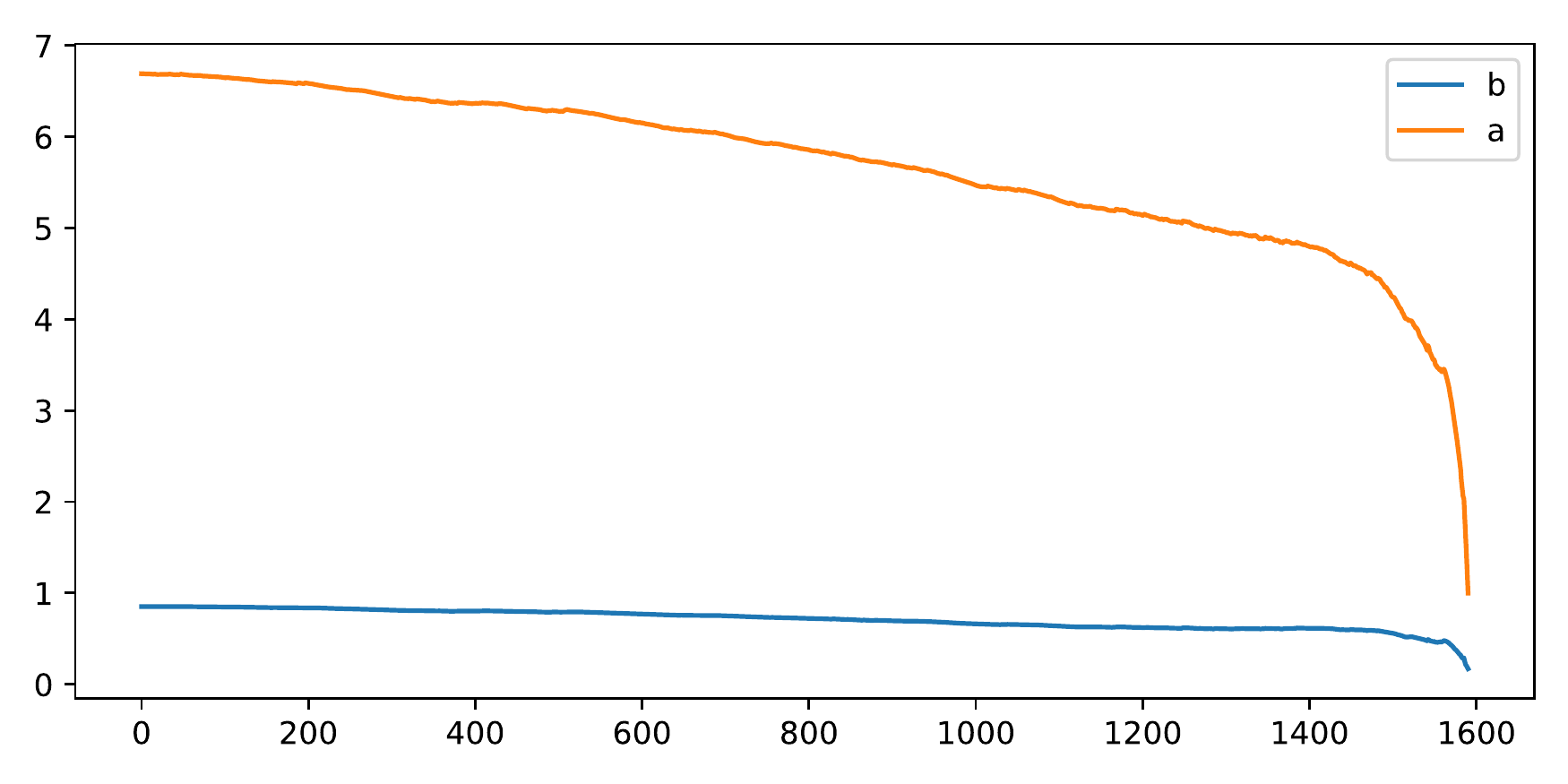}
	\caption{The following graph shows $a$ and $b$ values calculated from the Assam-Guwahati region Earthquake data. The $b$ value tends to be closer to $0.85$ indicating moderate-high number of small scale earthquakes. Total Mean Deviation of $b$ value is $~0.368$
}
	\label{fig:fig3}
\end{figure}
\begin{figure}
	\centering\includegraphics[scale=0.5]{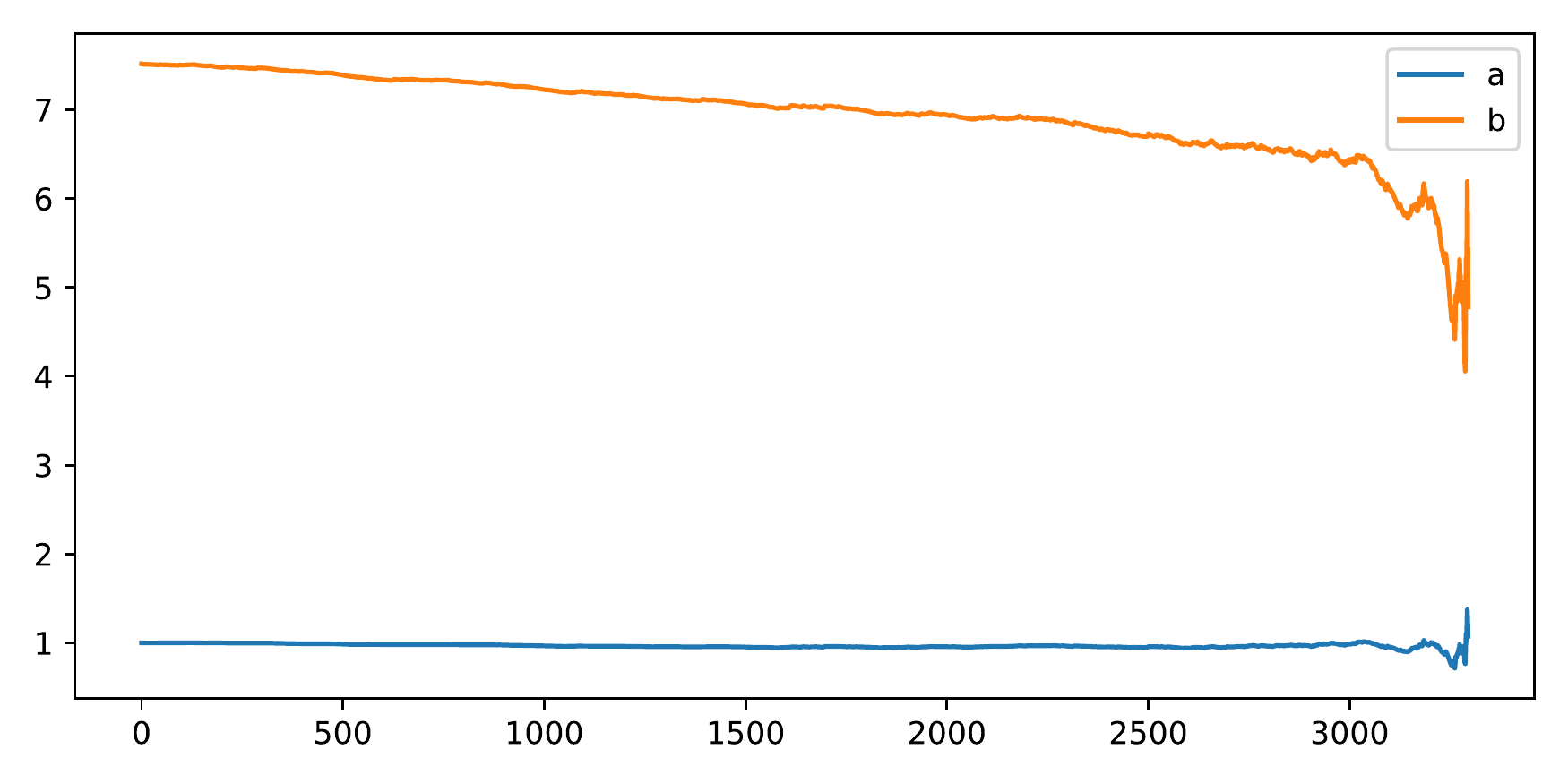}
	\caption{The following graph shows $a$ and $b$ values calculated from the California Earthquake data. The $b$ value is closer to $1$ indicating a seismically active region
}
	\label{fig:fig4}
\end{figure}

In earthquake prediction study for North-Eastern India, linear least square regression analysis based method is proposed along with real time calculation of $a$ and $b$ values in this study.

\subsection{Deviation from Gutenberg-Richter Law}
Deviation of actual data from Gutenberg-Richter inverse power law (\ref{gr_law}) is also considered as a seismic indicator \citep{serra}. We calculate it using the general variance model, where a greater $a-bM_i$ value corresponds to a greater conformance and therefore-- more likely to be predicted by the inverse power law:      \setcounter{equation}{8}
\begin{equation}
\sigma =\frac{\sum\left(\log N-a-bM\right)^2}{n-1}.
\end{equation}

\begin{figure}[H]
	\centering\includegraphics[scale=0.5]{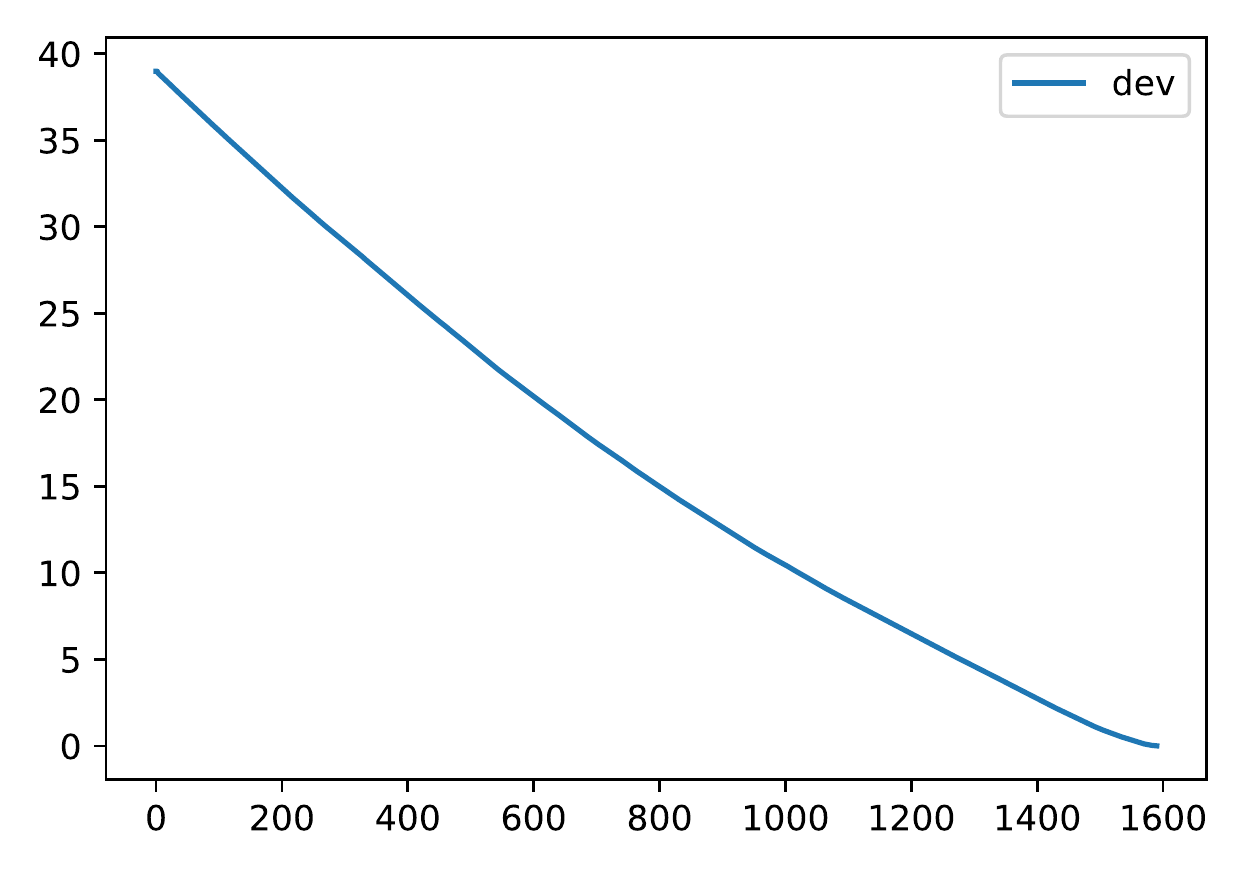}
	\caption{The graph shows the deviation from Gutenberg-richter law for the Assam-Guwahati region}
	\label{fig:fig7}
        \includegraphics[scale=0.5]{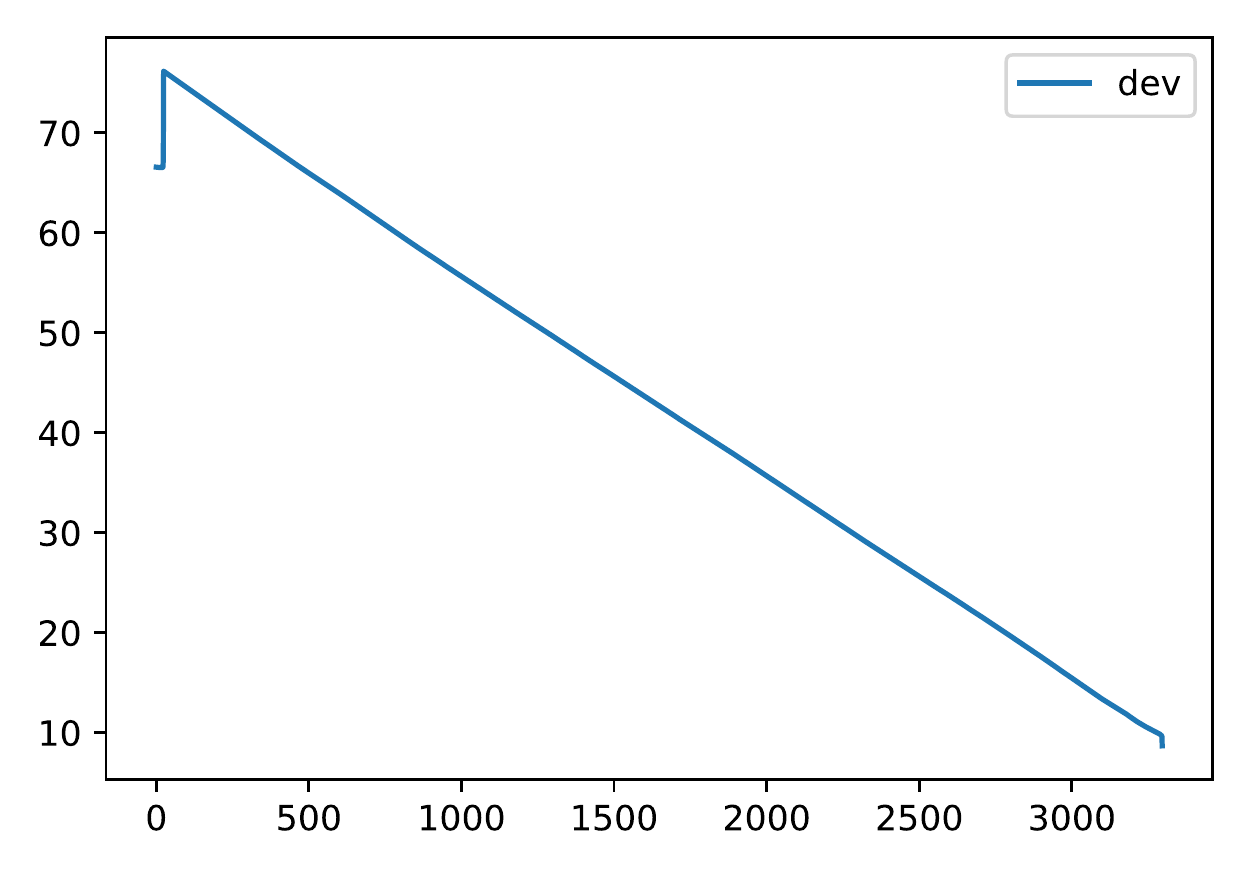}\caption{The graph shows the deviation from Gutenberg-richter law for the California region}
\end{figure}

\subsection{Expected Magnitude}
\begin{figure}
	\centering\includegraphics
	[scale=0.5]{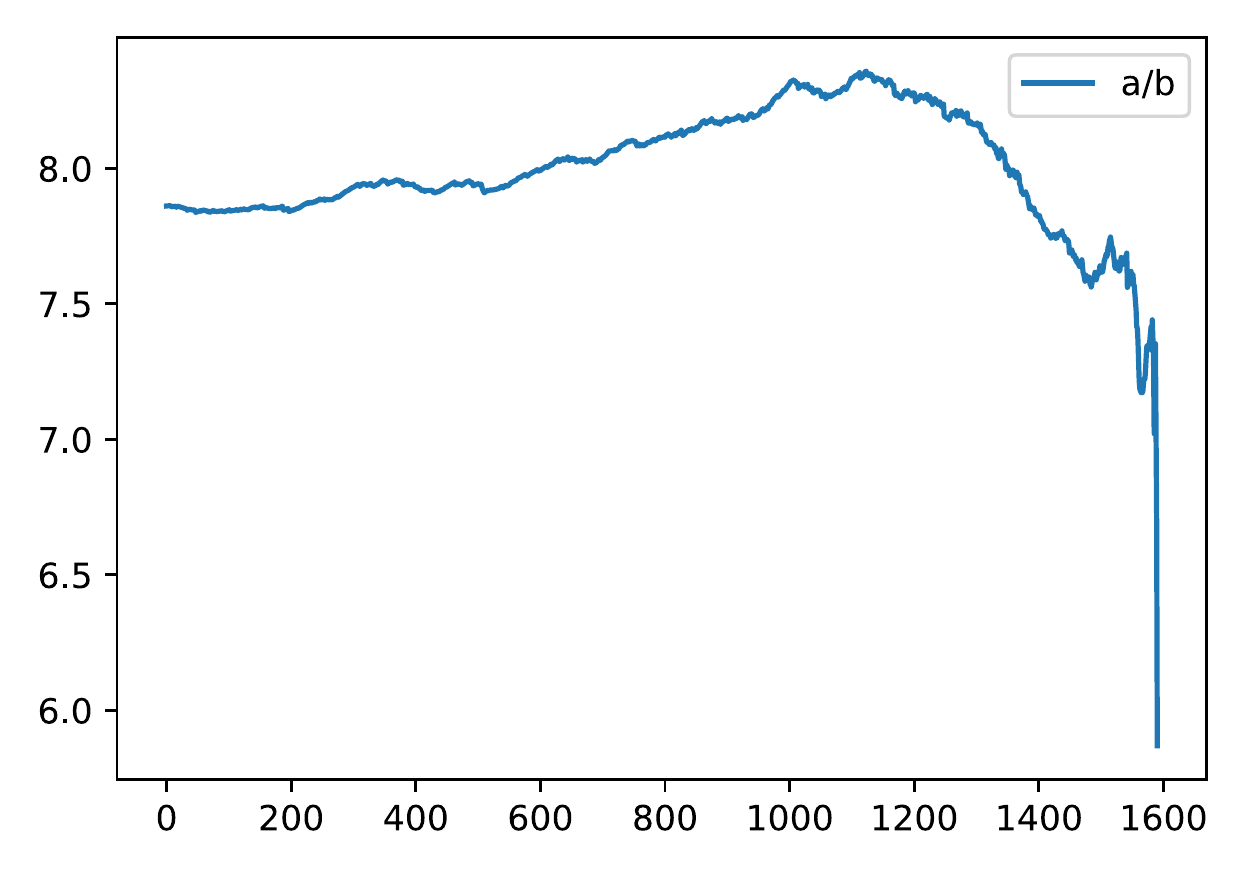}
	\caption{Expected Magnitude of the Assam-Guwahati region calculated as per the respective $a$ and $b$ values}
	\label{fig:fig8}
        \includegraphics[scale=0.5]{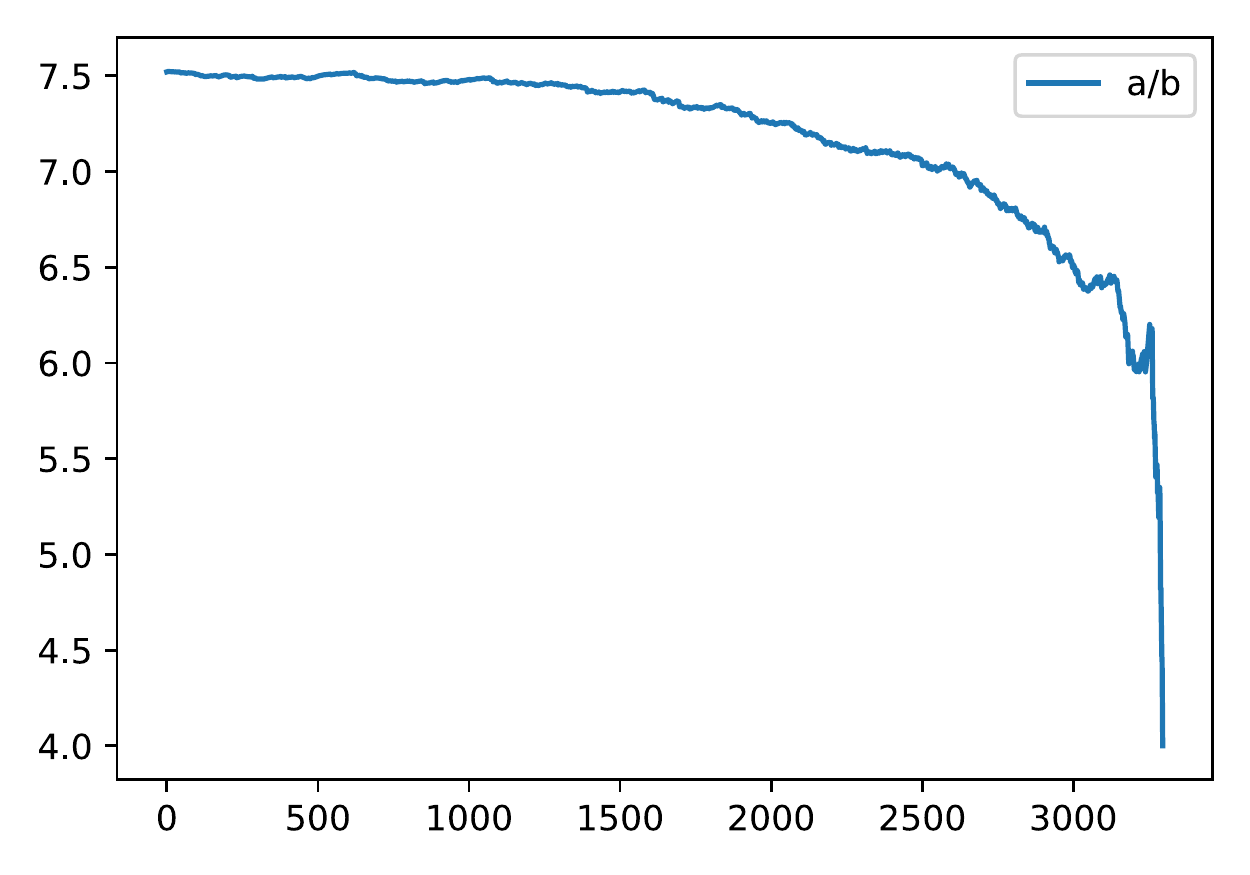}
        \caption{Expected Magnitude of the California region calculated as per the respective $a$ and $b$ values}
        \label{fig:fig9}
\end{figure}

The difference between the maximum observed and the maximum occurred earthquake magnitude is also considered as a seismic indicator (refer to figure \ref{fig:fig8} and \ref{fig:fig9}). The maximum observed event is listed in the catalog, while maximum expected event is obtained using the equation 
\begin{equation}
    M_\text{expected}=\frac{a}{b},
\end{equation}
where $a$ is the $y$-intercept in the inverse power law obtained from \ref{gr_law}.

\subsection{Maximum magnitude in last seven days}
The maximum magnitude recorded in the days previous to $E_t$ is also considered as an important seismic parameter (\citep{wang}; \citep{alarifi}) and is represented as
\begin{equation}
    x_{6i}=\max(M_i),\text{where }t\in[-7,0).
\end{equation}

\subsection{Total recurrence time}
It is also known as probabilistic recurrence time ($T_r$) and is defined as the time between two earthquakes of magnitude greater than or equal to $M'$ and is calculated using \ref{trt}. This parameter is another interpretation of Gutenberg-Richter’s law. As evident from the statement of inverse law, there will be different value of $T_r$ for every different value of $M'$, which would increase with increasing magnitude.
\begin{equation}
    T_r=\frac{T}{10^{a-bM'}}.\label{trt}
\end{equation}

Available literature does not focus on which value of $M'$ to be selected in such a scenario therefore $T_r$ is calculated for every $M'$ from $4.0$ to $6.0$ magnitudes following the principle of retaining maximum available information. So for two sets of $a$ and $b$ values along with varying $M'$ adds $42$ seismic features to the dataset.

\subsection{Probability of earthquake occurrence}
The probability of earthquake occurrence of magnitude greater than or equal to $6.0$ is also taken as an important seismic feature. It is represented by $x_{7i}$ and calculated through \ref{eo}. The inclusion of this feature supports the inclusion of Gutenberg-Richter law in an indirect way. The value of $x_{7i}$ is dependent upon the corresponding $b$ value:
\begin{equation}
    x_{7i}=e^{\frac{-3b_i}{\log e}}.\label{eo}
\end{equation}
Therefore, $b_\text{lsq}$ and $b_\text{mlk}$ are separately used to calculate $x_7i$, thus giving two different values for this seismic feature.

\section{Earthquake Magnitude Prediction model}
Unlike previous other earthquake magnitude prediction models proposed, in this paper a new learning algorithm was used which consists of Single Layer Feed-forward Neural network (SLFN). This algorithm along with minimum Redundancy Maximum Relevance (mRMR) added to the hardiness of the model. The layout of the final prediction is given below. The dataset of the region was divided into Training and Testing sets. $75\%$ of the data for training and testing was performed on the rest $25\%$ of the data. All the parameters were taken into consideration while training and their variance within time.

\begin{figure}[H]
	\centering\includegraphics[scale=.7]{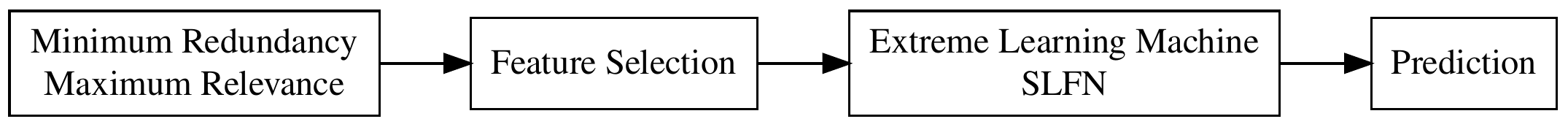}
	\caption{Flow Chart of the ELM Model used to predict earthquake magnitude in Assam and California}
	\label{fig:fig10}
 \end{figure}
 
The proposed procedure includes the two step feature selection. The features are selected after performing relevancy and redundancy checks, to make sure that only useful features are employed for earthquake prediction. The selected set of features was then passed on to Extreme Learning Machine (ELM).

\subsection{Extreme Learning Machine (ELM)}
ELM consists of a Single Layer Feed-Forward Neural Network where the parameters of the hidden nodes may not be tuned\citep{hanzhao2015}. This is an excellent feature which avoids the use of other Optimization Algorithms like Particle Swarm Optimization (PSO)\citep{kennedy}, Salp Swarm Optimization (SSA)\citep{ssa}. Here is a brief explanation of ELM \citep{huang}.

Given $N$ distinct training samples $(\mathcal{X}_i,\mathfrak{t}_i)$, where $\mathcal{X}_i=[x_{i1},x_{i2},\ldots,x_{in}]^T\in\mathbb{R}^n$ and $\mathfrak{t}_i=[t_{i1},t_{i2},\ldots,t_{in}]^T\in\mathbb{R}^m$, the output of a SLFN with $Y$ hidden nodes (additive or RBF nodes) can be represented by: \begin{equation}
o_j=\sum\limits_{i=1}^Y \beta_i f_i(\mathcal{X}_j)=\sum\limits_{i=1}^Y \beta_i g_i \left(\mathcal{W}_i\cdot\mathcal{X}_j+b_i\right),\hspace{1in}\text{for $j\in [1,N]$}.
\end{equation}where $\mathcal{W}_i=[w_{i1},w_{i2},\ldots,w_{in}]^T$ represents the weight vector that connects the hidden node and $j$th output. $\beta_i=[\beta_{i1},\beta_{i2}.\ldots,\beta_{im}]^T$ represents the weight vector that connects the hidden node and the $i$th output nodes.

$o_j$ is the output vector of the SLFN with respect to the input sample $x_i$. $a_i=\left[a_{i1},a_{i2},\ldots,a_{in}\right]^T$ and $b_i$ are learning parameters generated randomly of the $j$th hidden node, respectively.

The standard of SLFNs and L hidden nodes in the activation function $g(x)$ can be taken as samples of N without error. In other words,
\begin{equation}
    \sum\limits_{i=1}^Y \beta_i g_i\left(\mathcal{W}_i\cdot\mathcal{X}_j+b_i\right)=t_j,\hspace{1in}\text{for $j\in [1,N]$}.
\end{equation}

From the equations given for $N$, it can then be presented as \begin{equation}H\beta = T\label{eq:H=bt}\end{equation}
where 
\begin{equation}
    H=
    \begin{bmatrix}
    f(a_1x_1+b_1)&\cdots&f(a_Yx_1+b_Y)\\
    \vdots&\ddots&\vdots\\
    f(a_1x_N+b_1)&\cdots&f(a_Yx_Y+b_Y)\\
    \end{bmatrix}_{N\times N},
    \beta=\begin{bmatrix}
        \beta_1^T\\
        \vdots\\
        \beta_Y^T
    \end{bmatrix}_{Y\times m},
    T=\begin{bmatrix}
        t_1^T\\
        \vdots\\
        t_N^T
    \end{bmatrix}_{Y\times m}.
\end{equation}

To minimise the cost function $\lVert O-T\rVert$, ELM theories claim that the hidden nodes’ learning parameters $a$ and $b$ can be assigned randomly without considering the input data. Then, \ref{eq:H=bt} becomes a linear system and the output weights $\beta$ can be analytically determined by finding a least-square solution as follows, 
\begin{equation}
\beta-H^\dag T
\end{equation}
Where $H^\dag$ represents the Moore-Penrose generalized inverse of $H$. Hence, by mathematical transformation the output weights are calculated. This avoids the lengthy training phase where adjustment of the parameters takes place iteratively along with some learning parameters (such as learning rate and iterations). 

\tikzset{%
  every neuron/.style={
    circle,
    draw,
    minimum size=1cm
  },
  neuron missing/.style={
    draw=none, 
    scale=4,
    text height=0.333cm,
    execute at begin node=\color{black}$\vdots$
  },
}
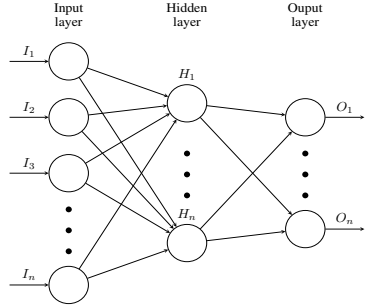
\begin{figure}[H]
\centering
\resizebox{2.1in}{1.6in}{
\begin{tikzpicture}[x=1.5cm, y=1.5cm, >=stealth]

\foreach \m/\l [count=\y] in {1,2,3,missing,4}
  \node [every neuron/.try, neuron \m/.try] (input-\m) at (0,2.5-\y) {};

\foreach \m [count=\y] in {1,missing,2}
  \node [every neuron/.try, neuron \m/.try ] (hidden-\m) at (2,2-\y*1.25) {};

\foreach \m [count=\y] in {1,missing,2}
  \node [every neuron/.try, neuron \m/.try ] (output-\m) at (4,1.5-\y) {};

\foreach \l [count=\i] in {1,2,3,n}
  \draw [<-] (input-\i) -- ++(-1,0)
    node [above, midway] {$I_\l$};

\foreach \l [count=\i] in {1,n}
  \node [above] at (hidden-\i.north) {$H_\l$};

\foreach \l [count=\i] in {1,n}
  \draw [->] (output-\i) -- ++(1,0)
    node [above, midway] {$O_\l$};

\foreach \i in {1,...,4}
  \foreach \j in {1,...,2}
    \draw [->] (input-\i) -- (hidden-\j);

\foreach \i in {1,...,2}
  \foreach \j in {1,...,2}
    \draw [->] (hidden-\i) -- (output-\j);

\foreach \l [count=\x from 0] in {Input, Hidden, Ouput}
  \node [align=center, above] at (\x*2,2) {\l \\ layer};

\end{tikzpicture}
}
\caption{A diagram showing the basic ELM Neural Network structure, with the input layer, a hidden layer and an output layer}
\end{figure}

\citep{huangl} listed the variables, where $H$ represents the output matrix of the hidden layer of the neural network. In $H$, the $i$th column is used to describe the $i$th hidden layer nodes in terms of the input nodes. If $L\leq N$ represents the desired number of hidden nodes, the activation function g becomes infinitely differentiable. 

\section{Results}

 The data in this study came from United States Geological Survey. The model was run on a Lenovo G-80 laptop with Intel® Core™ i$5-5200$U CPU @ $2.20$GHz and $8.00$ GB RAM. The results can be concluded into the following table:
The ELM Model got an RMSE of $~0.08$ which indicates over-fitting in the model. The average relative error percentage in other papers came out to be around 3\%. The SVR had an RMSE of $0.043$ with a heavy training time of $2289$ seconds. A significant drop in the testing time is observed in the ELM.

Using this model along with the CTS-M (Continuous Time Serious Markov) model which makes it possible to predict the location region of the next earthquake. This can improve the efficiency of the Early Warning Systems by a mile and also improve time efficacy of disaster response teams.
As ELM proved to be better than SVR, it was used to predict earthquake magnitude trained on California Earthquake Data and tested on the Assam-Guwahati Region.

\begin{figure}[H]
\centering\includegraphics[scale=0.6]{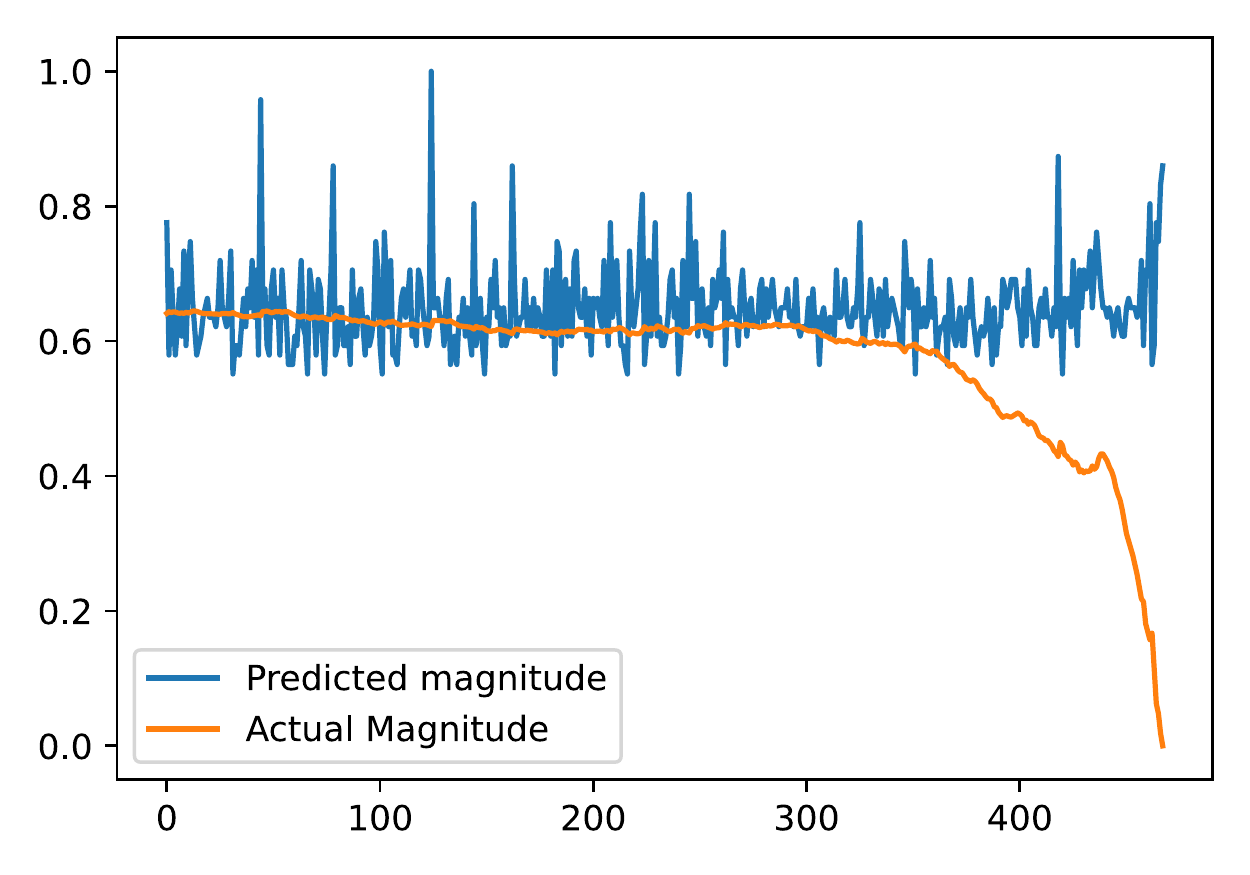}
\caption{following graph contains the scaled magnitude values predicted by the ELM Neural Network trained over Assam Region vs scaled actual magnitude values}
\end{figure}

\begin{figure}
	\centering
     \subfloat[\centering Overall RMSE]{{\includegraphics[scale=0.5]{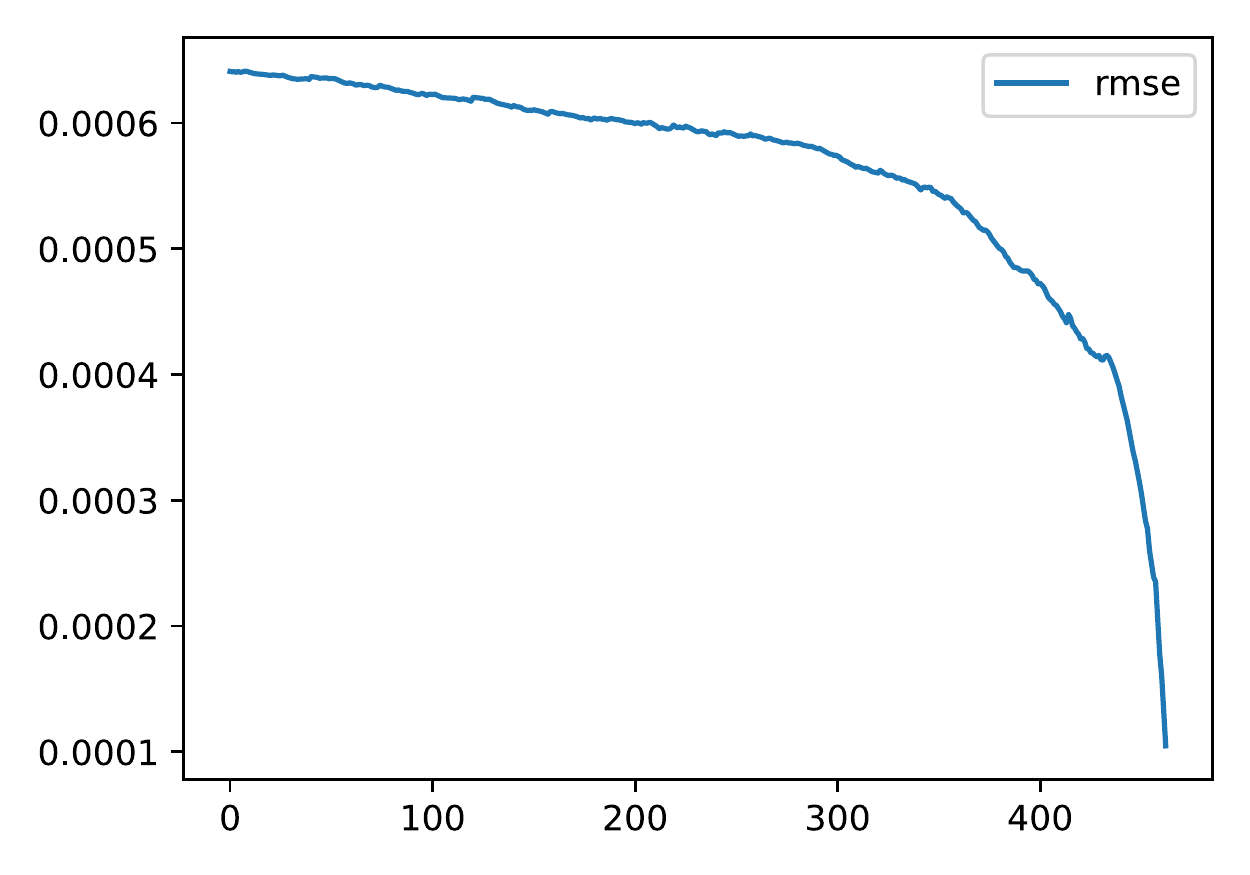} }}%
    \qquad
    \subfloat[\centering Parametric error (cumulative testing error)]{{\includegraphics[scale=0.5]{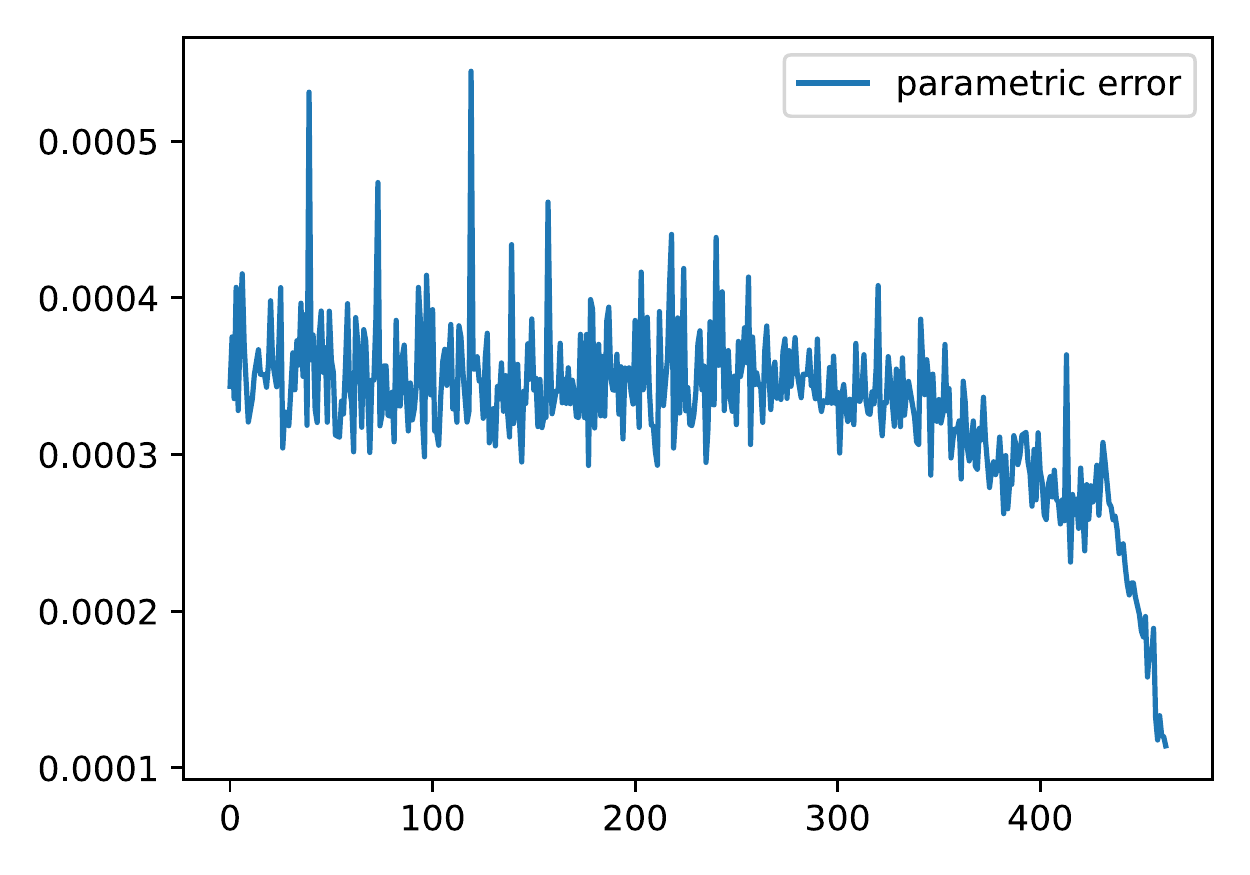} }}%
    \caption{RMSE and parametric error of all parameters for ELM trained and tested on the Assam-Guwahati region Data}%
    \label{fig:rmse}%
\end{figure}

All models were tested on the Assam-Guwahati region with ratio of $3:1$ with the training data. for the California dataset, it had $26192$ data points and was tested on $8639$ data points.

\section{Conclusion}
Two Machine Learning techniques have been used to predict earthquakes in the subducting Assam-Guwahati Region, one of the most seismically active regions in India. Both Magnitude Predictors shows significant results when compared to one another.

Further, the better Machine Learning Model (Extreme Learning Machine) was trained on California Dataset and tested on the same Assam-Guwahati Data. 
This can be improved by using an M-ELM model which reduces the chances of over-fitting. In the end, it can drastically improve the retro-traditional machine learning models. The parameter $b$ (commonly referred to as the "$b$-value") is commonly close to 1.0 in seismically active regions. This means that for a given frequency of magnitude $4.0$ or larger events there will be $10$ times as many magnitude $3.0$ or larger quakes and $100$ times as many magnitude $2.0$ or larger quakes. There is some variation of $b$-values in the approximate range of $0.5$ to $2$ depending on the source environment of the region. In a notable portion of the data, the $b$ value seems to less than $1$ and gradually decreases.

The study used real time $b$ value change across time ($b(t)$), change recorded after every earthquake occurrence, instead of taking the $b$ value to be constant throughout one year time. 
The study shows, although earthquake occurrence is supposed to be decidedly nonlinear and appears to be a random phenomenon, yet it can be modeled on the basis of geophysical facts of the seismic region along with highly sophisticated modeling and learning approaches of machine learning. 

\begin{table}
	\caption{Result of Earthquake models on the Assam-Guwahati region}
	\centering
	\begin{tabular}{llll}
		\toprule
		     & Assam-ELM     & SVR & Cali-ELM\\
		\midrule
		Testing Time & $0.109375$s  & $198$s &  $0.5$s \\
		RMSE     & $0.0081675$ & $0.043$ &$0.4572$    \\
		Training Time     & $3.140625$s       & $2289$s & $4$s \\
		\bottomrule
	\end{tabular}
	\label{tab:table1}
\end{table}

\begin{table}
	\caption{Minimum/Maximum and mean $a$ and $b$ values calculated on the Assam-Guwahati data}
	\centering
	\begin{tabular}{lll}
		\toprule
		     Min $b$-value & Max $b$-value & Mean $b$-value\\
		\midrule
		$0.17019$ & $0.8533$ & $0.3685$\\
	\end{tabular}\\
        \begin{tabular}{lll}
		\toprule
		     Min $a$-value & Max $a$-value & Mean $a$-value\\
		\midrule
		$0.9988$ & $6.6923$ & $5.6804$\\
		\bottomrule
	\end{tabular}
	\label{tab:table2}
\end{table}
\begin{itemize}
    \item ELM model- trained and tested on Assam region proves to be better.
    \item A Significant Drop ($99.95\%$ in $T_e$, ) in the Testing and Training Time can be observed in the California (ELM) and Assam (ELM) models $(T_e<1s, T_r<5s).$
    \item There exists a trade-off between $T_e$ and RMSE of ELM in Assam Data and California, indicating that the model is more sensitive to training data samples.
\end{itemize}

A smaller $b-$value likely suggests that the stress is high in the examined region. Decreasing $b$ value within the seismogenic volume under consideration has been found to correlate with increasing effective stress levels prior to major shocks. 

Normal faulting is associated with the highest $b-$values, strike-slip events show mean values and thrust events the lowest values. This observation means that $b$ acts as a stress meter, depending inversely on the differential stress.

\bibliographystyle{unsrtnat}
\bibliography{references}  %

\end{document}